%% file: arvix_zll.tex
\documentclass[conference,compsoc]{IEEEtran}
\usepackage{hyperref}
\usepackage[utf8]{inputenc}
\usepackage{amsmath}
\usepackage{graphicx}
\usepackage{fancyhdr}
\usepackage{url}
\usepackage{caption}
\usepackage{subcaption}
\usepackage{tabularx}
\usepackage{cleveref}
\newcolumntype{L}[1]{>{\raggedright\arraybackslash}p{#1}} % linksbündig mit Breitenangabe
\newcolumntype{C}[1]{>{\centering\arraybackslash}p{#1}} % zentriert mit Breitenangabe
\newcolumntype{R}[1]{>{\raggedleft\arraybackslash}p{#1}} % rechtsbündig mit Breitenangabe

\usepackage[usenames,dvipsnames]{xcolor}

\definecolor{darkspringgreen}{rgb}{0.09, 0.45, 0.27}

\newcommand{\paragraphsmall}[1]{~\newline \textit{#1} }

\ifCLASSOPTIONcompsoc
  \usepackage[nocompress]{cite}
\else
  \usepackage{cite}
\fi
\ifCLASSINFOpdf
\else
\fi
\IEEEoverridecommandlockouts

\begin{document}
\title{All Your Bulbs Are Belong to Us: \\Investigating the Current State of Security in Connected Lighting Systems}

\author{\IEEEauthorblockN{Philipp Morgner, Stephan Mattejat, Zinaida Benenson}
\IEEEauthorblockA{Friedrich-Alexander-Universität Erlangen-Nürnberg}
\thanks{A peer-reviewed revision of this work was published in the Proceedings of the 10th ACM Conference on Security \& Privacy in Wireless and Mobile Networks (WiSec'17): }
\thanks{{Philipp Morgner, Stephan Mattejat, Zinaida Benenson, Christian Müller, Frederik Armknecht -- Insecure to the Touch: Attacking ZigBee 3.0 via Touchlink Commissioning. ACM WiSec'17, Boston, MA, USA. }}
\thanks{DOI: 10.1145/3098243.3098254}
%\and
%\IEEEauthorblockN{Stephan Mattejat}
%\IEEEauthorblockA{University of Erlangen-Nuremberg}
%\and
%\IEEEauthorblockN{Zinaida Benenson}
%\IEEEauthorblockA{University of Erlangen-Nuremberg}
}


% compsoc conferences compared to traditional conferences), use this
%Telephone: (800) 555--1212, Fax: (888) 555--1212}


\maketitle

%%%%%%%%%%%%%%%%%%%%%%
% ABSTRACT
%%%%%%%%%%%%%%%%%%%%%%

\begin{abstract}
ZigBee Light Link (ZLL) is the low-power mesh network standard used by connected lighting systems,  such as Philips Hue, Osram Lightify, and GE Link.
These lighting systems are intended for residential use but also deployed in hotels, restaurants, and industrial buildings.
In this paper, we investigate the current state of security in ZLL-based connected lighting systems.
We extend the scope of known attacks  by describing novel attack procedures to show that the ZLL standard is insecure by design.
Using our penetration testing framework, we are able to take full control over all three systems mentioned above.
Besides novel attack procedures, we also extend the intended wireless range of max. 2 meters for configuring a ZLL device  to over 30 meters,
thus making ZLL-based systems susceptible to war driving. 
We conclude with a discussion about the security needs of connected lighting systems and derive several lessons for Internet of Things security that can be learned from the insecure design of ZLL-based connected lighting systems.
\end{abstract}

% \fi
\IEEEpeerreviewmaketitle

%%%%%%%%%%%%%%%%%%%%%%
% INTRODUCTION
%%%%%%%%%%%%%%%%%%%%%%

\input{s1_introduction}

%%%%%%%%%%%%%%%%%%%%%%
% CONNECTED LIGHTING SYSTEMS
%%%%%%%%%%%%%%%%%%%%%%

\input{s2_lighting_systems}

%%%%%%%%%%%%%%%%%%%%%%
% ZIGBEE
%%%%%%%%%%%%%%%%%%%%%%

\input{s3_zigbee}

%%%%%%%%%%%%%%%%%%%%%%
% RELATED WORK
%%%%%%%%%%%%%%%%%%%%%%

\input{s8_related_work}

%%%%%%%%%%%%%%%%%%%%%%
% THREAT MODEL
%%%%%%%%%%%%%%%%%%%%%%

\input{s4_threat_model}

%%%%%%%%%%%%%%%%%%%%%%
% SECURITY ANALYSIS
%%%%%%%%%%%%%%%%%%%%%%

\input{s5_security_analysis}

%%%%%%%%%%%%%%%%%%%%%%
% DISCUSSION
%%%%%%%%%%%%%%%%%%%%%%

\input{s6_discussion}

%%%%%%%%%%%%%%%%%%%%%%
% CONCLUSION
%%%%%%%%%%%%%%%%%%%%%%

\input{s9_conclusion}


\bibliographystyle{IEEEtran}
\bibliography{bib-zll-philipp,bib-zll-zina}

\end{document}

%% file: s1_introduction.tex
\section{Introduction}
\label{sec:intro}

The interest of consumers in connected lighting for residential environments increased rapidly since 2012.
The major reason is the introduction of new product lines of connected lighting systems that can be controlled via remote control, smartphone or tablet.

In September 2012, the LIFX bulb was presented on Kickstarter \cite{lifx}, an online crowdfunding platform, and raised 1.3 million dollar within two months. LIFX light bulbs connect to the home WiFi network, can be controlled via a mobile device and and require no additional hardware.
A few weeks later, Philips introduced Hue, which is considered the most popular connected lighting system for homes today \cite{hue1}. The Philips Hue system comprises white-color as well as RGB-color LED light bulbs, LED strips, wall and ceiling lights, as well as portable lights.
Philips Hue has an open API for developers to build third-party applications. We found\footnote{As of August 2, 2016} 123 Android apps in the Play Store and 96 iOS apps in the AppStore that relate to Philips Hue. 
Philips' consumer luminaries, the sector of which Philips Hue is the main product line, had a worldwide revenue of more than 500 million Euros in 2015\footnote{Consumer luminaries take 7\% of the overall sales of Philips lighting division according to Philips Annual Report 2015 \cite{hue2}.}.

In 2014, Osram and GE Lighting entered the market with their connected lighting systems called Lightify \cite{osram1} and Link \cite{ge1}, respectively.  
During the Eurovision Song Contest of 2015, an Osram Lightify system combined with a smartphone voting app reflected the atmosphere  in the public viewing area to ``give the major event a further emotional touch'' \cite{osram_esc}. %of the visitors
Osram Lightify offers white-color and RGB-color LED light bulbs, LED stripes, ceiling and wall lights, as well as garden spotlights, while GE Link supplies only white-color LED light bulbs focusing on the US market. All these lighting devices can be controlled via wall switches, smartphones or tablets.
Philips Hue, Osram Lightify, and GE Link are based on ZigBee Light Link (ZLL) standard,  
while the LIFX lighting system is based on the 6LoWPAN mesh network standard that is not subject of our investigation in this paper.

\begin{figure}[t]
	\centering
	\includegraphics[width=0.48\textwidth]{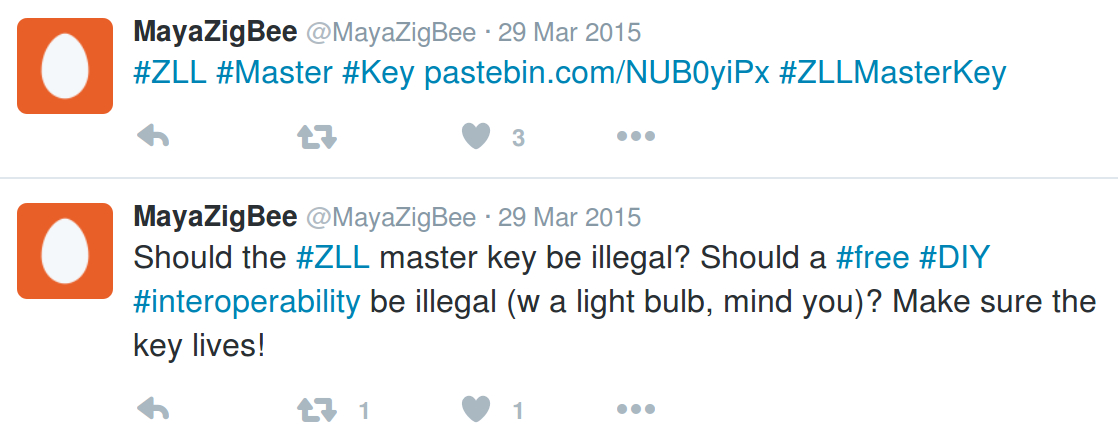}
	\caption{The ZLL master key was leaked via Twitter in March 2015.} 
	\label{fig:intro:twitter}
\end{figure}

The ZLL standard provides two procedures for setting up a new network for ZLL devices, or to bootstrap a new ZLL device to an existing ZLL network: classical commissioning and touchlink commissioning.
Although classical commission is usually applied, each ZLL-certified device must implement both commissioning procedures. 
The security features of the ZLL touchlink commissioning procedure rely on a global \emph{ZLL master key} that is used to encrypt the current network key before this key is transmitted to the joining device. This ZLL master key is distributed to manufacturers of ZLL-certified devices under a non-disclosure agreement (NDA). 
However, in March  2015, the ZLL master key was leaked on Twitter\footnote{https://twitter.com/mayazigbee} as shown in Figure \ref{fig:intro:twitter}. Ever since the leakage of the ZLL master key, the touchlink commissioning procedure is considered to be insecure.  %generally assumed

% Attacks: Twitter

\paragraphsmall{Goals.}
In this paper, we investigate the state of security in ZLL-based connected lighting systems.
To this end, we systematize the known attacks  that have been previously presented, mainly in a non-academic context. 
Furthermore, we provide a security analysis of the ZLL touchlink commissioning procedure which has not been 
part of a comprehensive security analysis before, to the best of our knowledge.
We provide full technical details and a comprehensive evaluation of the three most popular connected lighting systems that facilitate the ZLL standard: Philips Hue, Osram Lightify, and GE Link.
We further discuss which level of security is appropriate for connected lighting systems and which lessons can be learned for Internet of Things (IoT) security from the insecurity of ZLL.

\paragraphsmall{Results.}
Our evaluation shows that the classical commissioning procedure  is deployed per default by the manufacturers.  
Attacks against the classical commissioning procedure are known and successfully tested with the connected lighting systems Philips Hue and Osram Lightify\cite{sans, zillner2015, zillner2015deepsec}. 
We focus on the touchlink commissioning procedure, which also provides an attractive target since it must be implemented in each ZLL-certified device,
and describe novel attacks that  exploit design flaws in the touchlink specification of the ZLL standard.
We implement a penetration testing framework, evaluate these attacks, and describe difficulties we encountered during our investigations.
We are able to force bulbs to accept a new network key, which allows us to send commands to the bulbs, e.g., to turn them on or off, or to change the color.
In this analysis, we discovered further attacks that can be performed even without the knowledge of the ZLL master key indicating that these systems have been insecure even before the leakage of the ZLL master key.
For example, we are able to remove arbitrary light bulbs from the legitimate networks. 
We can also cause bulbs to blink for several hours, while the legitimate user has no chance to shut the bulbs down except through physical disconnection.

We prove that the three popular connected lighting systems Philips Hue, Osram Lightify, and GE Link are vulnerable to these attacks.
Besides novel attack procedures, we also increase the attack range significantly. 
In fact, touchlink commissioning is intended to work in close proximity (max. 2 meters) only. In our evaluation, we are able to control ZLL-certified devices using touchlink commands from a distance of 15 to 36 meters, depending on the manufacturer. 
Finally, we derive four critical points of failure in IoT security that can be learned from the insecure design of the ZLL standard.

\paragraphsmall{Implications.}
The security mechanisms of ZLL-based connected lighting systems contain fallback solutions and flaws 
that allow an attacker to remotely control the lights from a certain distance.
We show that these security weaknesses  have a real-world impact on three popular ZLL-based connected lighting systems.
Since ZLL devices can be remotely controlled from a distance of more than 30 meters, these devices are vulnerable to war driving and targeted attacks.
As connected lighting systems gain an increasing popularity since 2012, we hope that our findings can
raise the attention of the security community to improve security measures in connected lighting systems and other residential IoT networks.

\paragraphsmall{Outline.} 
The paper is structured as follows. We introduce the system architecture of ZLL-based connected lighting systems from the consumer perspective in \Cref{sec:systems}.
In \Cref{sec:zigbee}, we describe relevant technical details of the ZLL standard, and outline the ZLL security features in \Cref{sec:zllsec}.
We systematize previous attacks in  \Cref{sec:related}.
The threat model that we assume for the remaining part of this paper is presented in \Cref{sec:model}.
We perform the security analysis of touchlink commissioning, which contains the description of novel attacks as well as their evaluation, in \Cref{sec:analysis}.
The implications of our findings are discussed in \Cref{sec:discussion}.
This paper concludes in \Cref{sec:conclusion}.

%% file: s2_lighting_systems.tex
\section{ZLL-based Connected Lighting Systems}
\label{sec:systems}

We investigate the security of the three popular connected lighting systems by Philips, Osram and GE
that use the ZigBee Light Link standard as of May 2016. 
We introduce the consumer view on these systems in this section, and then present technical details of the ZLL standard in Section \ref{sec:zigbee}.

\subsection{System Architecture}
\label{sec:systems:arch}

% Figure: System Architecture
\begin{figure}[htb]
	\centering
	\includegraphics[width=0.48\textwidth]{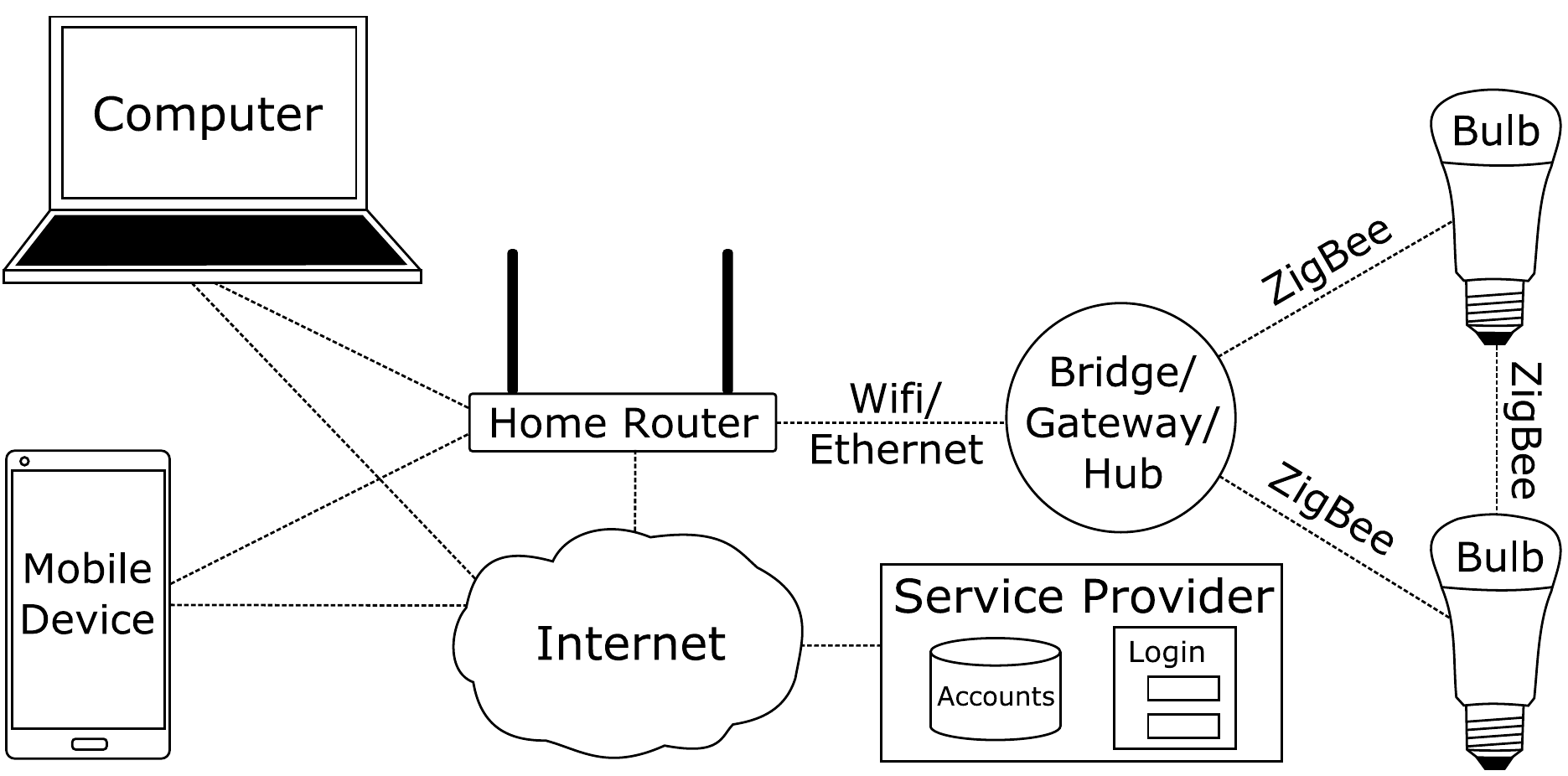}
	\caption{System architecture of a ZLL-based connected lighting system with two bulbs.}
	\label{fig:arch}
\end{figure}

\begin{figure*}[htb!]
\begin{subfigure}[c]{0.33\textwidth}
	\centering
	\includegraphics[height=1in]{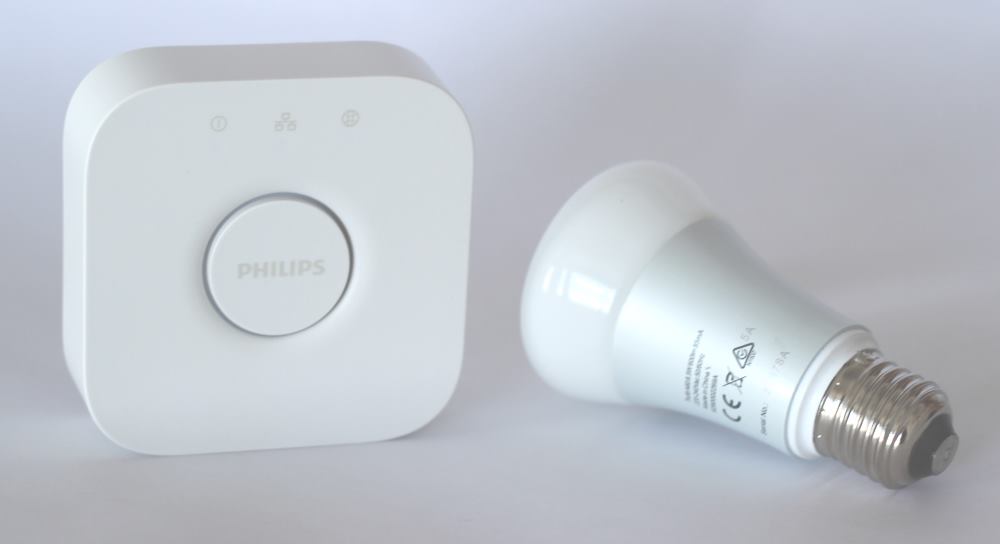}
	\caption{Philips Hue}
	\label{fig:hue}
\end{subfigure}
\begin{subfigure}[c]{0.33\textwidth}
	\centering
	\includegraphics[height=1in]{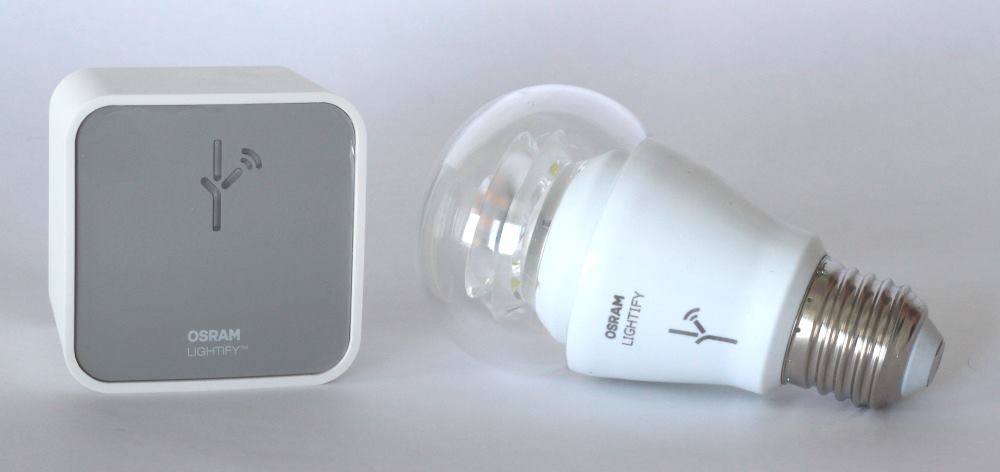}
	\caption{Osram Lightify}
	\label{fig:lightify}
\end{subfigure}
\begin{subfigure}[c]{0.33\textwidth}
	\centering
	\includegraphics[height=1in]{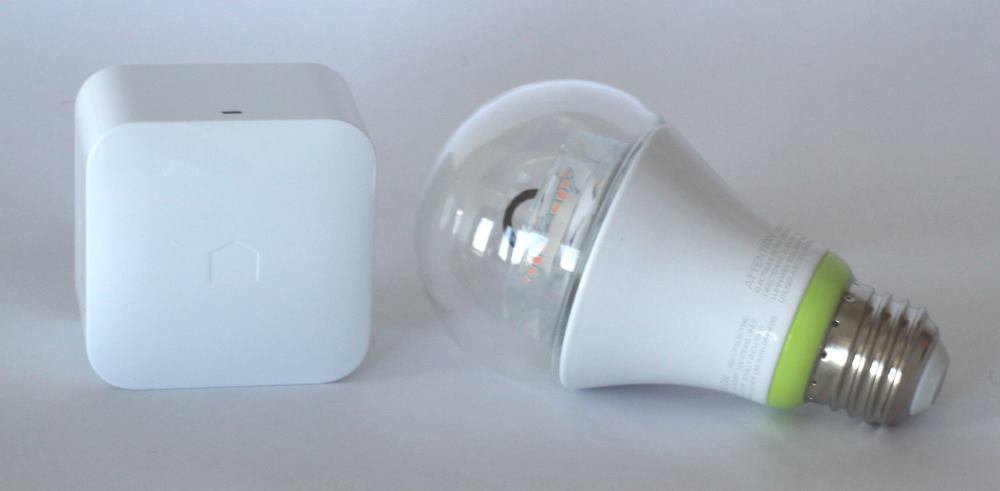}
	\caption{GE Link}
	\label{fig:link}
\end{subfigure}
\caption[Product pictures of the three evaluated connected lighting systems.]{Product pictures of the three evaluated connected lighting systems.}
\end{figure*}

The general system architecture is presented in \Cref{fig:arch}. 
Each connected lighting system consists of at least one smart light bulb and a bridge (also referred to as gateway or hub) that is used as a ZigBee transceiver to communicate with the bulbs. The bridge connects to the home router via Ethernet (Philips) or WiFi (GE, Osram).

Connected bulbs can be physically turned on and off using regular light switches, but this is usually not desired since off-state bulbs cannot be controlled by mobile devices anymore. 
To turn on or off as well as to change color and brightness of the light, a manufacturer- or third-party app on a mobile device (or a computer) is required. The user uses the app to send commands via the Internet and/or home router to the bridge, which translates the commands into ZigBee command frames and transmits them to the light bulbs.  There also exist remote controls (Philips) and ZigBee dimmer switches (Philips, Osram) that can directly send ZigBee commands to the bulbs without connecting to the home router. 

\subsection{System Setup}
\label{sec:systems:setup}

We analyzed a Philips Hue starter set including one bridge and three white and color ambiance LED bulbs, which is exemplary shown in Figure \ref{fig:hue}. 
Furthermore, we deployed an Osram Lightify gateway with a classic A60 tunable white LED bulb as second evaluated lighting system as shown in Figure \ref{fig:lightify}. 
Our third connected lighting system is the GE Link starter pack containing a Link hub and two Link A19 soft white LED light bulbs, similar to Figure \ref{fig:link}.

For the initial setup of each system, the user installs the manufacturer-specific app on a mobile device. In addition, it is mandatory to register a user account for the Osram Lightify or GE Link system. For Philips Hue, the registration of an online account is only mandatory if the user wants to remotely control the lights, e.g., from the work place. In the deployment of the Osram Lightify and GE Link systems, the user connects his or her mobile device to a WiFi-network that is provided by the bridge using the WiFi credentials printed at the back of the bridge. Then, the user is asked to enter the WiFi credentials of the home router into the app and as a result, the bridge connects to the home router.
To deploy the Philips Hue system, the users connects the bridge with the home router using an included Ethernet cable. 
Finally, the user commissions the bridge and the bulbs that belong to a single connected lighting system to a new ZLL network by following the instructions of the manufacturer-specific app.

%% file: s3_zigbee.tex
\section{ZigBee Light Link}
\label{sec:zigbee}

ZigBee is a wireless low-power standard that connects embedded technologies in personal area networks (PANs).
PANs are networks that allow the communication among personal devices such as computers, mobile phones, sensors, or household devices.
Compared to WiFi, ZigBee-certified devices send smaller packets and consume far less energy, while ZigBee has a larger wireless range than the regular Bluetooth.

\subsection{ZigBee Overview}
\label{sec:zigbee:overview}

The ZigBee specifications are maintained by the ZigBee Alliance, a global non-profit organization that comprises over 250 members. 
The ZigBee Alliance defines the network, security, and application layers and supervises the conformance and interoperability of ZigBee-certified products.
The ZigBee specifications include different \emph{application profiles} comprising customized sets of features and protocols 
for specific application areas, such as ZigBee Home Automation (ZHA) \cite{spec:zigbee:ha}, ZigBee Smart Energy (SE)  \cite{spec:zigbee:se}, or ZigBee Light Link (ZLL) \cite{spec:zigbee:ll}. 
Application profiles differ in functionality as well as in security levels. 

In this paper, we focus on the ZLL application profile, which was developed with the contribution of Philips, Osram, GE, STMicroelectronics, Greenwave, Ember (acquired by Silicon Labs in 2012), Texas Instruments, Atmel, and NXP \cite{zigbee-webinar}.

A new revision of the ZigBee specification, \textit{ZigBee~3.0}, was released in December 2015 \cite{zigbee3} to ZigBee Alliance members but has not been disclosed to the public yet. \textit{ZigBee~3.0} is announced to replace a subset of the existing application profiles (including ZLL) but also to provide backwards compatibility. 
As of the beginning of August 2016, there are no products available that are certified for \textit{ZigBee~3.0}.

\subsection{Stack Architecture}
\label{sec:zigbee:arch}

The ZLL stack consists of four layers: physical (PHY), medium access control (MAC), network (NWK), and application (APL).
The two lower layers, PHY and MAC, are defined in the IEEE 802.15.4-2003 specification \cite{spec:ieee:802:15:4:2003}. This specification is also incorporated into other WPAN standards, most prominently Thread Group \cite{webinar:thread} and WirelessHART \cite{spec:wirelesshart}.
The PHY layer is the lowest layer and defines the physical interface. ZigBee uses radio-frequency communication in the 2.4 GHz ISM band, which is divided into 16 channels of each utilizing 2 MHz bandwidth. 
The MAC layer provides functionalities to transmit data frames in a reliable manner by managing access to the radio channel via CSMA/CA mechanisms, sending beacon frames, acknowledgement frames, and performing synchronization techniques.

The NWK layer is specified in the 2012-ratified \textit{ZigBee Pro} \cite{spec:zigbee:pro} standard, and manages network topology, routing, security services, and acts as a message broker.
The APL layer implements the application profiles, which is, in the case of connected lighting systems, the ZLL profile.

\section{ZLL Security}
\label{sec:zllsec}

\subsection{Security Goals}
\label{sec:zigbee:goals}

% Security Goals
While ZLL standard does not define security goals \cite{armknecht}, the ZigBee Pro specification describes security assumptions \cite[p. 426]{spec:zigbee:pro} but no security goals either. 
For this reason, we define following security goal that applies to ZLL networks: 
Only the legitimate user should be able to add, control, or remove devices in a ZLL network, i.e., the legitimate user always keeps control over all devices in a ZLL network.

\subsection{Security Mechanisms}
\label{sec:zigbee:encryption}

% Encryption
Security measures in ZLL are only applied to the NWK layer. The MAC layer, as defined in the IEEE~802.15.4 standard, supports different encryption and authentication mechanisms. However, these security mechanisms are not implemented in ZLL-certified devices. Also, the ZLL profile does not support security at APL layer.

ZLL-certified devices facilitate the AES-CCM* authenticated encryption scheme\footnote{Compared to AES-CCM (without asterisk), this specific mode allows also encryption-only or integrity-only variants.} using a $128$-bit network key, which is shared between all devices of a network to secure the communication.
During the initial setup, a ZLL device performs a commissioning procedure to obtain the network key.

\subsection{Commissioning}
\label{sec:zigbee:comm}

Commissioning is the process in which a new ZLL network is set up or a new ZLL device is added to an existing network. 
The ZLL standard specifies two commissioning procedures: classical commissioning and touchlink commissioning.
Each ZLL-certified device supports both commissioning procedures, which differ in key management schemes and key transport protocols.

\subsubsection{Classical Commissioning}
\label{sec:zigbee:comm:classical}

The classical commissioning procedure is specified in the ZigBee Pro standard and is supported by ZigBee devices of different application profiles, e.g., ZLL or ZHA.
One purpose of the classical commissioning procedure is to connect ZigBee devices of different application profiles.

The network discovery is triggered by the device that wants to join a network by sending \textit{beacon requests} on different channels.
If a coordinator of an existing ZLL network is open for new devices, this coordinator replies with a \textit{beacon response} containing information about the network including a flag indicating joining permission. 
Then the new ZLL device decides whether it joins the network or not.

In contrast to the classical commissioning in regular ZigBee networks where a trust center handles the key management, there is no trust center deployed in ZLL networks.
Therefore, the coordinator device sends the network key directly to the joining device. This network key is encrypted using a ZLL-specific global \emph{ZLL link key}, which is distributed to the manufacturers under an NDA. The ZLL link key has not been leaked to the best of our knowledge.

Security weaknesses in the classical commission procedure, especially a fallback mechanism to the publicly known \emph{global default trust center link key}, are described in \Cref{sec:related}. 

\subsubsection{Touchlink Commissioning}
\label{sec:zigbee:comm:touchlink}

The touchlink commissioning procedure was firstly introduced in the ZLL standard, and is not supported by devices of other ZigBee application profiles.
Touchlink was especially designed to meet the usage requirements of connected lighting systems.
Compared to classical commissioning, touchlink commissioning provides an extended functionality that goes beyond the plain joining of devices.
One of the goals is to enable use cases in which the commissioning procedure is performed between low-function device, e.g., a remote control, and a bulb. For such scenarios, touchlink commissioning offers also the possibility to manage network features with the so-called touchlink commands.

\begin{figure}[h]
	\centering
	\includegraphics[width=.48\textwidth]{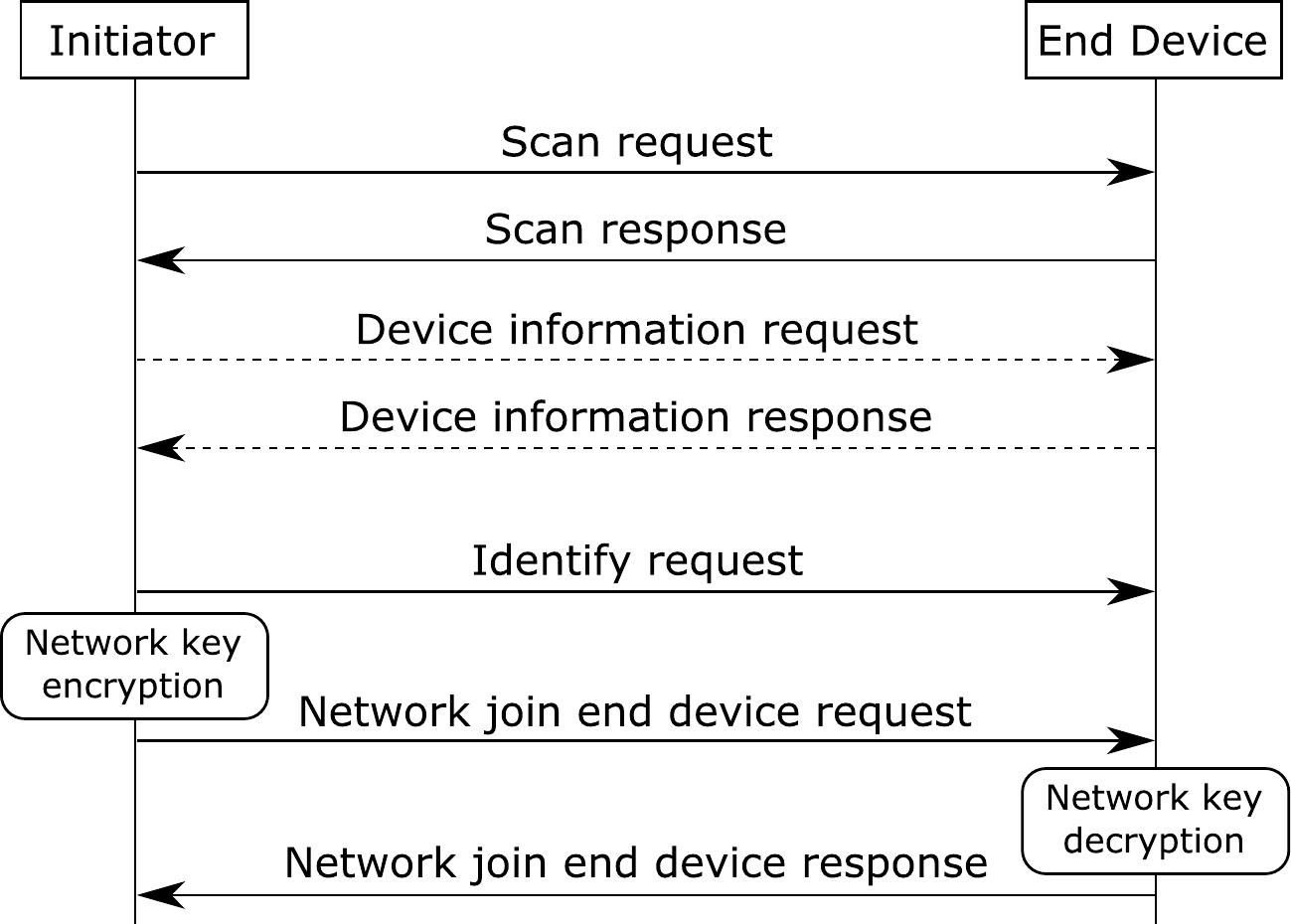}
	\caption{Touchlink commissioning protocol in the ZLL profile.} 
	\label{fig:zigbee:comm:ll}
\end{figure}

In Figure \ref{fig:zigbee:comm:ll}, we describe the commissioning protocol for joining a new device to an existing ZLL network. % in an technical manner, omitting details % Figure: Touchlink protocol
%
% ZLL Spec., p. 89f
The touchlink commissioning protocol is executed between two entities, an initiator and an end device. The initiator is usually either a controller, such as a remote control, or a bridge, i.\,e., a ZLL device characterized by a physical button which is pushed to start the commissioning procedure. 
The initiator is typically already a member of a network or is used to establish a new network. The end device represents the ZLL device that shall be added to the network, usually a LED light bulb.

First, the initiator starts the device scan procedure by sending \textit{scan requests} on different channels as defined in the ZLL specification. These scan requests include a randomly generated transaction identifier. 
The end device replies with a \textit{scan response} containing the same transaction identifier, a random response identifier, the list of all supported  keys, and further information.
On receipt of a \textit{scan response} from an end device, the initiator may request more information  about the sub-devices of the target in case there exists multiple sub-devices, e.g., if the light bulb is equipped with multiple lights that can be controlled separately. In this case, the initiator sends a \textit{device information request} and the targeted end device responses with the \textit{device information response}.  

The device scan may yield multiple potential devices from which the user can select one for the next steps.
The user has the option to send an \textit{identify request} to a device, upon which the target device performs a predefined identification action, e.g., a light bulb will flash  for a few seconds. An \textit{identify request} contains the corresponding transaction identifier as well as the duration of the identification action.

To join a new end device to a network, the initiator encrypts the current network key using the procedure described in Section \ref{sec:zigbee:comm:key}, builds a \textit{network join end device request} containing the encrypted network key, transaction identifier, the key index as well as further network information, and then sends this command frame to the selected end device. 
On receiving the message, the joining end device replies with a \textit{network join end device response} indicating success.

\subsubsection{Network Key Encryption}
\label{sec:zigbee:comm:key}

During the touchlink commissioning procedure, the initiator encrypts the network key and sends this encrypted key to the joining end device. 

The \textit{scan response} indicates the keys available to the end device, in our scenarios the \emph{ZLL master key}. 
The \emph{ZLL master key} is distributed to certified manufacturers of ZLL products and bound with an NDA.

To encrypt the randomly chosen network key, the initiator expands the transaction identifier  and the response identifier  to a $128$-bit number, which represents the plaintext input of the AES-ECB encryption, while the \emph{ZLL master key} is used as encryption key. The resulting ciphertext output is called \emph{transport key}. In the next step, the actual network key is encrypted using the AES-ECB encryption mode and the \emph{transport key}.

%% file: s8_related_work.tex
\section{Related Work}
\label{sec:related}

\subsection{Security in ZigBee and IEEE 802.15.4}
\label{sec:related:zigbee}

The security of the ZigBee standard as well as the underlying IEEE 802.15.4 standard attracted much less attention in the academic research community compared to other wireless standards, such as WiFi, Bluetooth or mobile telephony. 

Sastry and Wagner \cite{DBLP:conf/ws/SastryW04} analyzed the security mechanisms of the IEEE 802.15.4 protocol. However, these
mechanisms are not used in ZigBee, as we describe in \Cref{sec:zigbee:encryption}. 
 Wright \cite{wright2009} published the penetration testing tool KillerBee, which allows to sniff and analyze traffic of ZigBee and other IEEE 802.15.4-based networks. Wright also exposed that the network key of the then-current ZigBee standard was sent in clear text over the air. He demonstrated successful replay attacks using prior captured ZigBee traffic. The ZigBee Pro specification, released in 2012, addressed these security weaknesses.
Goodspeed et. al. \cite{DBLP:conf/hicss/GoodspeedBMSS12} developed exploration tools to analyze the wireless attack surface of IEEE 802.15.4 networks. 
Armknecht et. al. \cite{armknecht} present a formal security model for the ZLL touchlink commissioning.
Further papers \cite{DBLP:conf/his/OlawumiHAVT14, DBLP:conf/hicss/VidgrenHPRT13} cover security issues of ZigBee networks but these papers refer to security weaknesses concerning outdated ZigBee specifications and have not been evaluated with ZigBee-certified products.

\subsection{Attacks Against Connected Lighting Systems}
\label{sec:related:attacks}

Since the emergence of connected lighting systems in 2012, these systems have been subject to a number of security investigations.

Dhanjani \cite{dhanjani2013} published implementation weaknesses of the command authentication in the Philips Hue lighting system. He discovered that the secret whitelist token, which is required to authenticate the commands sent via the REST API from the app (or website) to the bridge, is an MD5 hash of the MAC address of the controlling device. 
He developed malware that computes whitelist tokens and permanently issues `all lights off' commands in an infinite loop to local Philips Hue bridge, since the IP addresses of the bridges are predictable. This results in local blackouts of all Philips Hue systems that are connected to the same LAN. 
In addition, Dhanjani describes threats concerning remotely issued blackouts based on password leaks. 

Chapman \cite{chapman2014} obtained the firmware of LIFX light bulbs via a JTAG debugger and extracted cryptographic key material through reverse engineering of the firmware. He used the knowledge of the keys to decrypt captured IEEE 802.15.4 packets from the communication between LIFX bulbs as well as to send arbitrary commands to the bulbs, e.g., to turn off the lights. Note that the LIFX system is not based on the ZLL standard.

Heiland \cite{rapid7} exposed vulnerabilities in the Osram Lightify system. 
Through reverse-engineering, he discovered that WiFi credentials are stored in plaintext in the iOS Lightify Home app. 
Also, he revealed that the communication between app and gateway is insufficiently authenticated.
Concerning the implementation of the ZLL standard, he criticized the lack of a routinely renewal of the network key.

Ronen and Shamir \cite{DBLP:conf/eurosp/RonenS16} used the Philips Lux lighting system, which is the white-color variant of Hue, to build a covert channel for the exfiltration of data from an isolated environment. 
Also, Ronen et. al. \cite{ronen} demonstrated an attack in which they switch Philips Hue bulbs on and off from a distance of over 70 meters. However, at the time of writing
no technical details were known due to the responsible disclosure procedure the authors of the attack initiated.

Zillner et al. \cite{zillner2015, zillner2015wp} exposed security weaknesses in the ZigBee specification regarding the classical commissioning procedure. 
They showed that ZLL lighting systems use publicly known fallback keys in the classical commissioning procedure for the initial key exchange, which allows the extraction of the network key assuming the key transport frame was captured by the attacker during the commissioning of a device to a network. In their scenario, the user can be tricked into recommission a device to a network via jamming or sending a `reset to factory default' command.
With the knowledge of the current network key, the attacker can join the network and send commands to other devices in the network.
In a demonstration, they showed that they can turn the lights of a Philips Hue bulb on and off \cite{zillner2015deepsec}. 
In another attack by Zillner et al., the attacker `steals' bulbs from a legitimate network and joins them to its own network through sending a `reset to factory default' command and waiting until the bulb searches for a new network. Since the bulb connects automatically to the first available network, no interaction of the user is required. 
The attack descriptions do not provide technical details, e.g., how to send a `reset to factory default' command. 
Their attacks have been proven to be successful from a distance of at 3 meters \cite{sans} using the RaspBee platform as radio transceiver.

In \Cref{sec:analysis}, we present novel attacks focusing on the touchlink commissioning procedure and give full technical details and evaluation results. 

%% file: s4_threat_model.tex
\section{Threat Model}
\label{sec:model}

The threat model is determined as follows:
The user of the ZLL network is trusted and honest, and installs the ZLL network as demanded by the manufacturer. 
The online account credentials for the remote access to the ZLL network are not disclosed.
The ZLL devices are certified by the ZigBee Alliance and  follow the protocols described in the ZigBee Pro and the ZLL specifications. 
Therefore, the ZLL devices are equipped with the actual ZLL master key.

The goal of the attacker is undermine the security goals described in \Cref{sec:zigbee:goals}, i.e., to take complete control of devices in the ZLL network.
Therefore, either the attackers' equipment needs to be within the wireless range of the targeted ZLL network, or alternatively, a remotely controlled radio transceiver is located in the range of the ZLL network. An example for a remotely controlled radio transceiver can be an IEEE 802.15.4 radio transceiver that has a GSM interface with a SIM card module. In this scenario, the attacker can trigger an attack remotely via sending an SMS to the device.
We assume that attackers are able to eavesdrop and inject packets in the wireless communication of  at least one ZLL device of the targeted ZLL network. 
We further assume that attackers have neither physical access to the ZLL devices nor to any interface of the local area network (LAN) or wireless LAN (WLAN) to which a ZLL gateway might be connected.

%% file: s5_security_analysis.tex
\section{Security Analysis of Touchlink Commissioning}
\label{sec:analysis}

\begin{figure}[thb]
	\centering
	\includegraphics[width=3.2in]{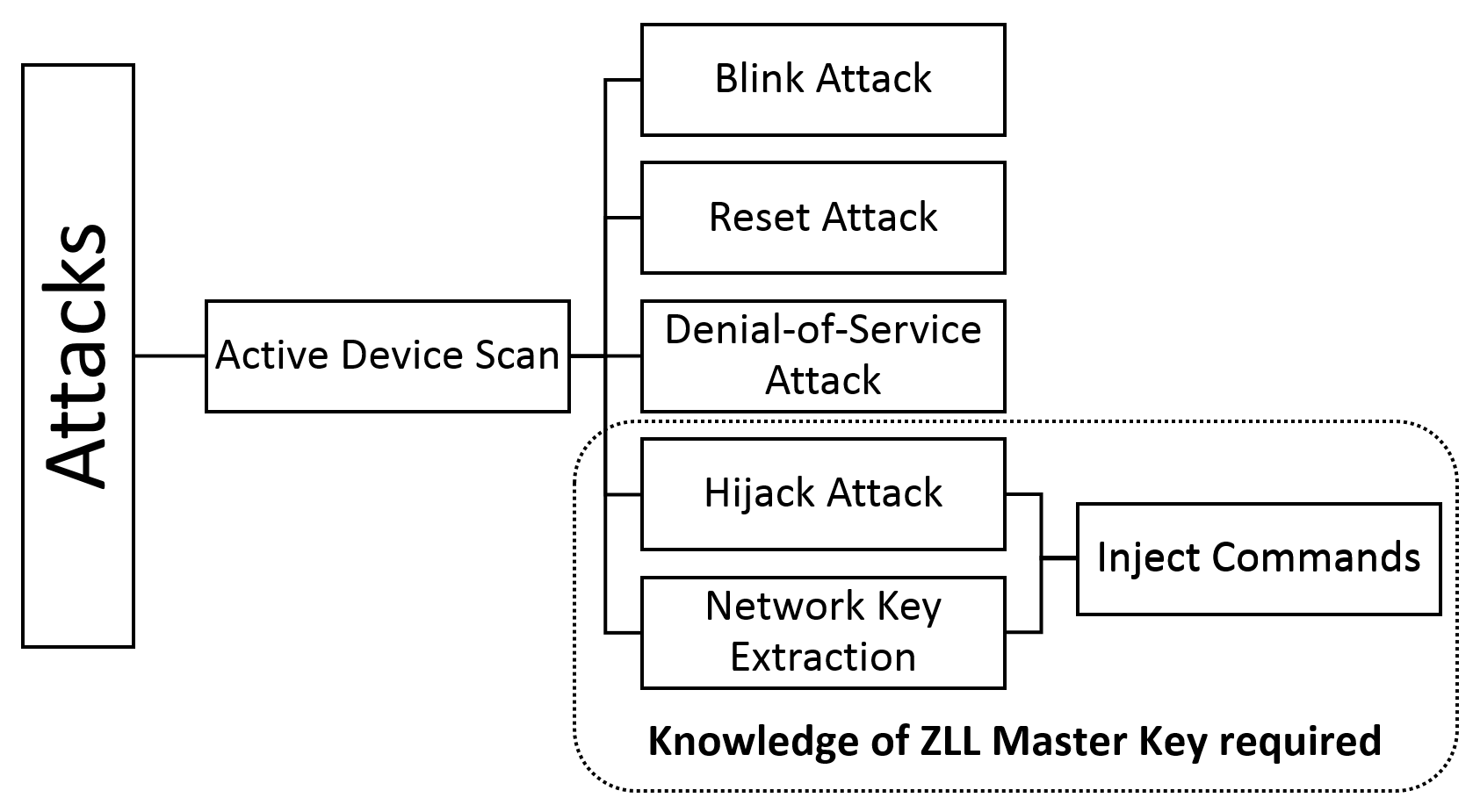}
	\caption{Overview of attacks.} 
	\label{fig:attacks:overview}
\end{figure}

We divide our attacks in two categories: In the first category, we describe attacks that exploit security weaknesses in the concept of so-called inter-PAN frames. These attacks require no knowledge of any cryptographic material. In the second category, we show attacks that need knowledge of the ZLL master key. Using this key, we gain control over multiple ZLL devices in a ZigBee network. In Figure \ref{fig:attacks:overview}, we provide an overview of all attacks that are described in this section. 

% Attacks: Overview

The attacks in this section outline the procedures to compromise a single ZLL device, which is in our scenario a connected light bulb. 
All attacks can be easily extended to target multiple ZLL devices at the same time by running the attack procedures parallel for different target devices.

During the evaluation, we realized that the connected lighting systems are not performing the touchlink commissioning procedure by default. Instead, the classical commissioning procedure described in the ZigBee Pro specification is deployed. Only the Hue lighting system provides an API command to trigger the touchlink commissioning procedure.
Nevertheless, all tested devices support the ZLL touchlink commissioning and are vulnerable to the presented attacks, even if they do not use this commissioning procedure by default.

%%%%%%%%%%%%%%%%%%%%%%%%%%%%%%%%%%%%%%%%%%%%%%%%%%%%%%%%%%%%%
%%%%%%%%%%%%%%%%%%%%%%%%%%%%%%%%%%%%%%%%%%%%%%%%%%%%%%%%%%%%%

\subsection{Penetration Testing Framework}
\label{sec:analysis:methodology}

For our research, we developed a penetration testing framework in Python to evaluate the security of ZLL devices.
This framework consists of three major components: 
First, a \emph{touchlink library} to build arbitrary touchlink packets and to keep track of source addresses and sequence numbers. 
Second, a \emph{crypto module} that provides the functionality to encrypt and decrypt ZigBee packets. This component also handles key transport frames, especially decrypting the encrypted network key, and vice versa.
Third, the \emph{radio interface module} enables the communication between the radio transceivers and the touchlink library.

We utilize two different hardware platforms  to send and receive ZigBee packets: the Moteiv Tmote Sky, a small sensor node, and the software-defined radio Ettus USRP B200.
The Tmote Sky features a MSP-430 micro controller and an IEEE 802.15.4-complaint radio. 
For the communication between the Tmote Sky and the touchlink library, we use the KillerBee framework \cite{wright2009}. 
The second radio transceiver, the USRP B200 from Ettus, is a software-defined radio covering the radio-frequency range between 70 MHz and 6 GHz.
The USRP features an FPGA and connects to a host computer via USB 3.0.
We use Scapy-radio \cite{scapyradio} as interface to send and receive ZigBee packets with the USRP. Scapy-radio itself uses capabilities of GnuRadio
and an IEEE 802.15.4 GnuRadio flow chart implementation~\cite{bloessl2013gnu}.

In addition, we implemented a command line tool that performs the attack procedures described in the following sections, and an injection module that is able to inject arbitrary packets.

%%%%%%%%%%%%%%%%%%%%%%%%%%%%%%%%%%%%%%%%%%%%%%%%%%%%%%%%%%%%%
%%%%%%%%%%%%%%%%%%%%%%%%%%%%%%%%%%%%%%%%%%%%%%%%%%%%%%%%%%%%%

\subsection{Testbed}
\label{sec:analysis:testbed}

In the setup of our evaluation, we follow the instructions of the corresponding manufacturers for the consumers, as described in \Cref{sec:systems}, to establish a realistic deployment simulating the conditions of a regular smart home. This indicates that all presented attacks apply to all ZLL networks, even if this network was originally set up using the classical commissioning procedure. %The setup of the connected lighting systems is described in \Cref{sec:systems}. 

Before starting our evaluation, we updated the Philips Hue firmware to the then-latest version 01031131 as well as the API to version 1.12.0.
We updated the Osram Lightify gateway WLAN to version 1.1.2.101 and the gateway ZigBee to version 1.2.0.67.
We found no possibility to update the GE Link firmware by using the manufacturer-recommended Wink app. 

The attacker equipment comprises a laptop on which our penetration testing framework, described in Section \ref{sec:analysis:methodology}, is installed. A radio transceiver, which is either the Tmote Sky or the Ettus USRP, is connected to the laptop. 
We started the evaluation of each attack with this default deployment, in which the lighting system works as intended and the system is not compromised.

%%%%%%%%%%%%%%%%%%%%%%%%%%%%%%%%%%%%%%%%%%%%%%%%%%%%%%%%%%%%%
%%%%%%%%%%%%%%%%%%%%%%%%%%%%%%%%%%%%%%%%%%%%%%%%%%%%%%%%%%%%%
\subsection{Exploitation of  Inter-PAN Frames}
\label{sec:analysis:interpan}

Inter-PAN frames are a special type of ZigBee frames that allow the communication between different personal area networks (PANs).
In the touchlink commissioning procedure, inter-PAN frames are used to transmit touchlink commands and their responses between initiator and end device.

Since there exists no shared key material between different PANs, inter-PAN frames are neither secured nor authenticated. 
Therefore, all attacks presented in the section are performed assuming no knowledge of the ZLL master key nor of any other cryptographic material relating to this devices.

%%%%%%%%%%%%%%%%%%%%%%%%%%%%%%%%%%%%%%%%%%%%%%%%%%%%%%%%%%%%%
\subsubsection{Active Device Scan}
\label{sec:analysis:interpan:scan}

%% PROCEDURE
The active device scan searches for ZLL devices in wireless range of the attackers equipment. The active device scan is a prior step that is mandatory in preparation of any further attack.

\paragraphsmall{Procedure.}
The attacker builds a \emph{scan request},
then sends this inter-PAN command frame on all ZigBee channels consecutively and listens a few milliseconds on each channel for \emph{scan responses}. 
The \emph{scan response} of each responding ZLL device contains the 32-bit response identifier, a list of all supported keys, and information about sub-devices as well as  other network-related information and parameters. 
Through the reception of \emph{scan responses}, the attacker learns about all ZLL devices that are also listening on this channel. 

ZigBee uses 16 channels in the 2.4GHz ISM band: channel 11 to 26, while channel 1 to 10 are located in other ISM bands. The ZLL profile specifications define four primary channels on which ZLL devices are listening for radio-frequency signals: 11, 15, 20, and 25. These channels are used for commissioning and normal operations, while all remaining channels can be used as backup.

%% EVALUATION
\paragraphsmall{Evaluation.}
The active network scan works with all three lighting systems. In general, all light bulbs responded to the \emph{scan request}, while also the Lightify gateway answers each time. The Link hub does not respond to \emph{scan requests}, and the Hue bridge only replies if the button on the hub was pushed within the last 30 seconds.
The state and color of the light bulbs is not changed, and also the connection between the legitimate controller and the bulbs is still working.

\paragraphsmall{ACK Spoofing.}
% Figure: Active Device Scan
\begin{figure}[htb]
            \centering
            \includegraphics[width=0.45\textwidth]{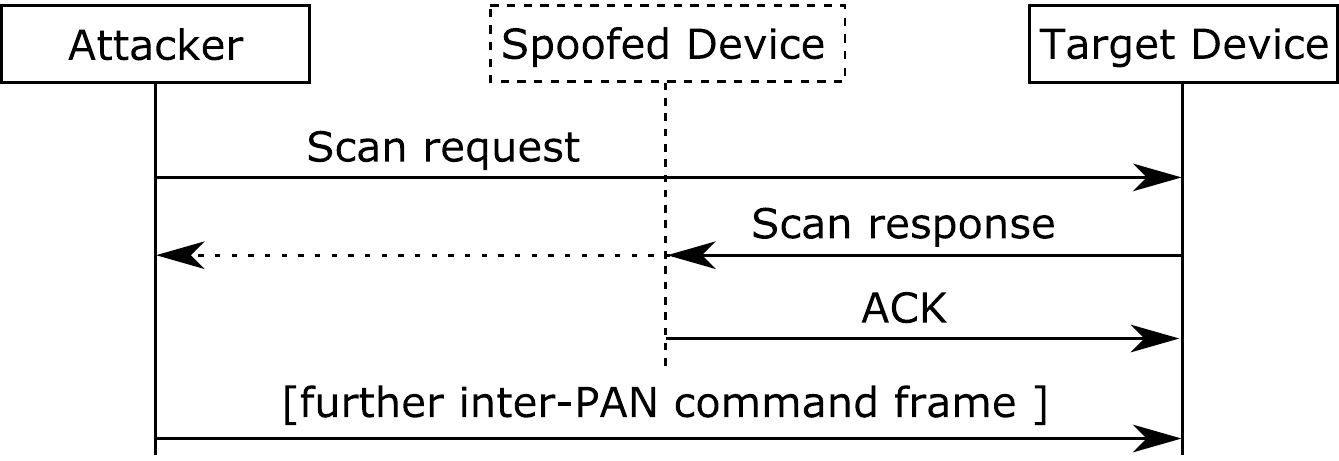}
            \caption[]%
            {{Acknowledgment spoofing.}}    
            \label{fig:attacks:ack}
\end{figure}

All following attacks start with an active network scan and then send further inter-PAN command frames. 
The reason for spoofing is that the Lightify bulb as well as the Link bulb demand a MAC-layer acknowledgment about the successful reception of the \emph{scan response} at the destination device. Otherwise, no further inter-PAN command frames would be accepted.
Because our implementation failed to send the acknowledgment within the demanded time frame of 864 microseconds, we spoof another ZigBee device in the network that acknowledges the reception of the \emph{scan response}, even if this device did not send the \emph{scan request}, as shown in \Cref{fig:attacks:ack}. With this trick, we are able to perform further attacks procedures (e.g., blink attack, reset attack, etc.) against Lightify and Link bulbs.
The spoofed device can belong to any arbitrary ZigBee network but has to listen on the same channel as the targeted device, and needs to be located in the wireless range of the target.
To spoof a device, we set the extended source address of the \emph{scan request} to the extended address of the spoofed device. 

In contrast, the Hue bulb responses to any arbitrary originator because apparently no acknowledgment on MAC-layer is required.

%%%%%%%%%%%%%%%%%%%%%%%%%%%%%%%%%%%%%%%%%%%%%%%%%%%%%%%%%%%%%
\subsubsection{Blink Attack}
\label{sec:analysis:interpan:blink}

%% PROCEDURE
The touchlink commissioning procedure provides the possibility to request a ZLL device to identify itself. 
Originally, the identify procedure is intended to give the resident a possibility to select and identify a certain bulb, which shall be added to the network.
From a technical perspective, the initiator sends an \emph{identify request} inter-PAN command frame to a selected end device, which starts flashing for a specified period of time.
This command can be abused by the attackers to annoy or frighten residents. 

\paragraphsmall{Procedure.}
After an active network scan and the reception the \emph{scan responses}, the attacker can send an \emph{identify request} to the targeted ZLL device. 
The \emph{identify request} contains the fields transaction identifier and identify duration.
The transaction identifier needs to be set to the same value as in the \emph{scan request}.
The identify duration can be at maximum 0xFFFE, which converts to a time duration of 18 hours 12 minutes 14 seconds. If the identify duration is set to 0, a previously started identify procedure is aborted before the specified duration elapsed. Setting the identify duration to 0xFFFF requests the bulb to perform the identify procedure for a default period of time, usually a few seconds.

At the reception of the \emph{identify request}, the targeted ZLL device starts blinking for the defined period of time. 

%% EVALUATION
\paragraphsmall{Evaluation.}
Each of the three presented lighting systems is vulnerable to the blink attack.
During the blinking of the lights, the users can neither turn off nor control the light bulb using apps provided by the manufacturer. 
The only way to shut down the lights is to physically disconnect the bulb from the power supply.
The attacker can abort the attack anytime by sending another \emph{identify request} with the field duration set to zero.
The maximum duration of blinking that can be triggered with a single \emph{identify request} is shown in Table \ref{tab:attacks:blink_duration}. 
We assume that the duration depends on the manufacturers implementation of the ZLL standard.

% Table: Maximum duration
\begin{table}
\center
\renewcommand{\arraystretch}{1.2}
\begin{tabular}{l r}
\cline{1-2} 
\textbf{System} & \textbf{Attack runtime}  \\ 
\cline{1-2}
Philips Hue &  18:12:14h \\
Osram Lightify & 9:12:53h \\
GE Link & 9:06:31h \\
\cline{1-2}
\end{tabular}
  \captionof{table}{The maximum duration of the blink attack.}
  \label{tab:attacks:blink_duration}
\end{table}

After performing the blink attack, the Hue bulb returns to the pre-attack state and color, while the Lightify bulb and the Link bulb change to the default state and color.
This attack also works if the bulb is originally turned off but supplied with power.

%%%%%%%%%%%%%%%%%%%%%%%%%%%%%%%%%%%%%%%%%%%%%%%%%%%%%%%%%%%%%
\subsubsection{Reset Attack}
\label{sec:analysis:interpan:reset}

%% PROCEDURE
In this attack, the attacker resets all settings of a ZLL device to the factory-new state. 

\paragraphsmall{Procedure.}
The attack is performed by sending a \emph{reset to factory new request}  inter-PAN command frame after a prior active network scan. 
The payload of the \emph{reset to factory new request} only contains the transaction identifier.
On the reception of a valid \emph{reset to factory new request}, the light bulb discards the current configuration. The color and brightness of the light bulb changes to the default states.
The user needs to recommission this device into the home network.

%% EVALUATION
\paragraphsmall{Evaluation.}
Our evaluation showed that all three lighting systems are vulnerable to the reset attack.
The color of the lights changes to the default color, which is in the most cases warm white.
Internally, all configuration parameter of the bulb are set to factory-new.

Because the bulb is factory-new, a legitimate user can easily reintegrate the bulb in the home network by searching for new devices in the app. This operation has to be initiated manually by the user.
In the mean time, i.e., before the user initiates a recommissioning, an attacker has the chance to hijack the resetted device using classical commissioning in combination with the publicly known fallback trust center link key as already demonstrated \cite{zillner2015}.

Interestingly, we are also able to reset the Lightify gateway (at any time) as well as the Hue bridge (if the button of the bridge was pushed within the last 30 seconds) to a factory-new state.

%%%%%%%%%%%%%%%%%%%%%%%%%%%%%%%%%%%%%%%%%%%%%%%%%%%%%%%%%%%%%
\subsubsection{Denial-of-Service Attacks}
\label{sec:analysis:interpan:dos}

%% PROCEDURE
We present two approaches to perform a denial-of-service attack:
In the first approach, we force a targeted ZLL device to change the current channel to another channel determined by the attacker.
In the second approach, we join the targeted device to a non-existent network.
In both cases, the user loses control over this device.
In our evaluation, denial-of-service attacks differ from the reset attack in the process of recovery: after a reset attack, the user can simply recommission the attacked bulb to the network again.  In the aftermath of a denial-of-service attack, the user needs to reset the bulb first before a recommissioning process to the network is possible. 

\paragraphsmall{Procedure.}
A change of the wireless channel can be enforced by sending a \emph{network update request}  inter-PAN command frame. The command must include a network update identifier that is higher than the current  update identifier of the targeted network, which  is a counter that is incremented every time when network settings are changed.
The current network update identifier can be retrieved from the \emph{scan response} of the target device, besides other network information.
After receiving the \emph{network update request}, which includes the new channel, the target device switches to this channel.
The legitimate network does not recognize the shift. 
As a consequence, the targeted device is not receiving user  commands anymore.

Using a \emph{network join end device request} inter-PAN command frame (instead of the \emph{network update request}), an attacker can manipulate additional network settings like the PAN ID and the current network key.
The attacker  sets the encrypted network key to a random 128-bit value.
The attacker sends the \emph{network join end device request} to the targeted ZLL device on the channel on which the device is currently listening. 
On the reception of a valid \emph{network join end device request}, the ZLL device leaves its current network and sets the internal parameters according to the new configuration.
The transaction is confirmed by sending a \emph{network join end device response}. The targeted ZLL device is not commissioned to the legitimate network anymore.

%% EVALUATION
\paragraphsmall{Evaluation.}
In the evaluation, all three presented lighting systems are vulnerable to both denial-of-service attacks.
The attack does neither change the color nor the state of the bulb but after performing the attack procedure, the targeted bulbs cannot be controlled by the user anymore.

Since the ZLL master key has been leaked, we can enhance the denial-of-service attack to the so-called hijack attack described in Section \ref{sec:analysis:masterkey:hijack}.

\paragraphsmall{Recovery.}
All lighting systems integrated functions to regain control over attacked devices.
However, these procedures are not obvious at  first sight or require technical knowledge.
In any case, a recovery entails manual effort for the user.

For GE Link and Osram Lightify, the only way to recover the attacked bulbs is a physical reset.
The physical reset is not specified in the ZLL standard, but can be achieved by powering the bulbs on and off in a certain manufacturer-specific pattern (e.g., 3s on, 5s off, repeat five times). This is a challenging task and might not always work the first time.

The Philips Hue system lacks a physical reset, to the best of our knowledge. However, the Hue system supports an additional commissioning mechanism \emph{manual search}, which is not specified in the ZLL standard. \emph{Manual search} works by entering a code that is printed on a Hue bulb into the Hue app. 
The \emph{manual search} fails if the channel of the attacked device was altered.

Touchlink commissioning can be applied as an alternative recovery procedure for Hue bulbs. It can be performed by using either the Hue API debug tool or a third-party app.
After the recommissioning with touchlink, an interesting effect can be observed:
Instead of reintegrating the attacked bulb into the former network, the Hue bridge detects that the network update identifier of the discovered device is higher than its own.
The Hue bridge adapts to the `latest' network settings and switches to the attacker-defined channel.
Consequently, the bridge loses the connection to all other bulbs, which remain on the former channel.
This behavior is specified in the ZLL specification.
Afterwards, all other bulbs of the former network have to be recommissioned to the new network using touchlink. 
This is a quite cumbersome task, because all devices have to be moved in close proximity (1-2 meters) to the Hue bridge in order to perform the touchlink commissioning.

%%%%%%%%%%%%%%%%%%%%%%%%%%%%%%%%%%%%%%%%%%%%%%%%%%%%%%%%%%%%%
\subsection{Exploitation of the Leaked ZLL Master Key}
\label{sec:analysis:masterkey}

%%%%%%%%%%%%%%%%%%%%%%%%%%%%%%%%%%%%%%%%%%%%%%%%%%%%%%%%%%%%%
\subsubsection{Hijack Attack}
\label{sec:analysis:masterkey:hijack}

The hijack attack extends the denial-of-service attack.  Instead of sending arbitrary bytes as the encrypted network key, the attacker forces the ZLL device to update its current network key to an attacker-chosen key.

%% PROCEDURE
\paragraphsmall{Procedure.}
Again, the attacker builds the \emph{network join end device request}  inter-PAN command frame as described in \Cref{sec:analysis:interpan:dos}.
The attacker-chosen network key is encrypted using the leaked ZLL master key, the transaction identifier from the \emph{scan request} and the response identifier from the scan response of the targeted ZLL device. This encrypted network key is included into a \emph{network update request} and sent to the targeted ZLL device.
On the reception of a valid \emph{network join end device request}, the ZLL device updates its internal parameters according to the received values and confirms the transaction by sending a \emph{network join end device response}. The targeted ZLL device is now commissioned to the network of the attacker, which has full control over this device.
The  legitimate network does not recognize the shift. As a consequence, commands send by the user will not be received anymore at the targeted ZLL device.

%% EVALUATION
\paragraphsmall{Evaluation.}
In the evaluation, we were able to force ZLL devices of all three connected lighting systems to accept an attacker-chosen key. This attack paves the way to send further application-specific commands to the ZLL devices.

%%%%%%%%%%%%%%%%%%%%%%%%%%%%%%%%%%%%%%%%%%%%%%%%%%%%%%%%%%%%%

\subsubsection{Network Key Extraction Attack}
\label{sec:analysis:masterkey:extraction}

%% PROCEDURE
An attacker is able to extract the current network key by eavesdropping the \emph{scan request}, \emph{scan response}, and the \emph{network join end device request} of an initial touchlink commissioning. Instead of capturing the \emph{network join end device request}, this attack can also be performed by capturing a \emph{network start request}. All these command frames must belong to the same transaction, i.e., contain the same transaction identifier. 

\paragraphsmall{Procedure.}
The user might be motivated to perform a touchlink commissioning procedure by the run of a prior reset or DoS attack (see Section \ref{sec:analysis:interpan:reset}).
After extracting the encrypted network key from the  \emph{network join end device request}, the network key is decrypted as described in Section \ref{sec:zigbee:comm:key}. The response identifier is known from the \emph{scan response}, while the transaction identifier is included in all packets belonging to the same transaction. 

%% EVALUATION
\paragraphsmall{Evaluation.}
For this attack, the user of a connected lighting system has to perform the touchlink commissioning procedure.  
Our investigations concluded that only Hue lighting systems can be targeted with this attack since only a few Hue third-party apps as well as the Hue remote control triggers this commissioning procedure.
To the best of our knowledge, there exist neither apps nor ZLL-certified devices for users of Lightify or Link to perform the touchlink commissioning procedure. 

In our evaluation setup, we sent a Hue API command to start a new touchlink commissioning procedure in an infinite loop and by pushing the button with an automatic construction, we were able to extract more than 13,000 valid network keys. All extracted keys seem to be unique. 
With the network key, an attacker can sniff and decrypt every packet being send in the network. Furthermore, the injection of commands is possible.

%%%%%%%%%%%%%%%%%%%%%%%%%%%%%%%%%%%%%%%%%%%%%%%%%%%%%%%%%%%%%
\subsubsection{Inject Commands}
\label{sec:analysis:masterkey:inject}

%% PROCEDURE
To send commands to the targeted device, we assume the knowledge of the current network key (which should not be confused with the leaked ZLL master key). The attacker knows the network key through performing either a prior hijack attack (see Section \ref{sec:analysis:masterkey:hijack}) or a prior network key extraction attack (see Section \ref{sec:analysis:masterkey:extraction}).

\paragraphsmall{Procedure.}
Assuming the attacker knows the current network key, the attacker can send commands to the ZLL devices. 
To control the bulbs, the attacker does not necessarily need to know the so-called destination endpoint of the lighting application of the bulb, since there is a broadcast endpoint with which all applications in a device can be addressed. 
The different commands and clusters to control the color and brightness of the lights are described in the ZigBee Cluster Library \cite{spec:zigbee:ll}.

%% EVALUATION
\paragraphsmall{Evaluation.}
Our evaluation of the command injection showed that all three lighting systems are vulnerable to this attack.
In our evaluation, we were able to send commands to turn the bulbs on and off and to change the light color of the Hue bulbs to any arbitrary color.
If the command injection follows a prior network key extraction attack, then the user is usually able to send own commands to bulbs along with the attacker. Otherwise, if the command injection follows a prior hijack attack, then the user has no possibility to regain control over the bulb except by performing the previously presented measures.

%%%%%%%%%%%%%%%%%%%%%%%%%%%%%%%%%%%%%%%%%%%%%%%%%%%%%%%%%%%%%
%%%%%%%%%%%%%%%%%%%%%%%%%%%%%%%%%%%%%%%%%%%%%%%%%%%%%%%%%%%%%

\subsection{Evaluation of Wireless Range}
\label{sec:analysis:range}

\begin{figure}
	\centering
	\includegraphics[width=0.48\textwidth]{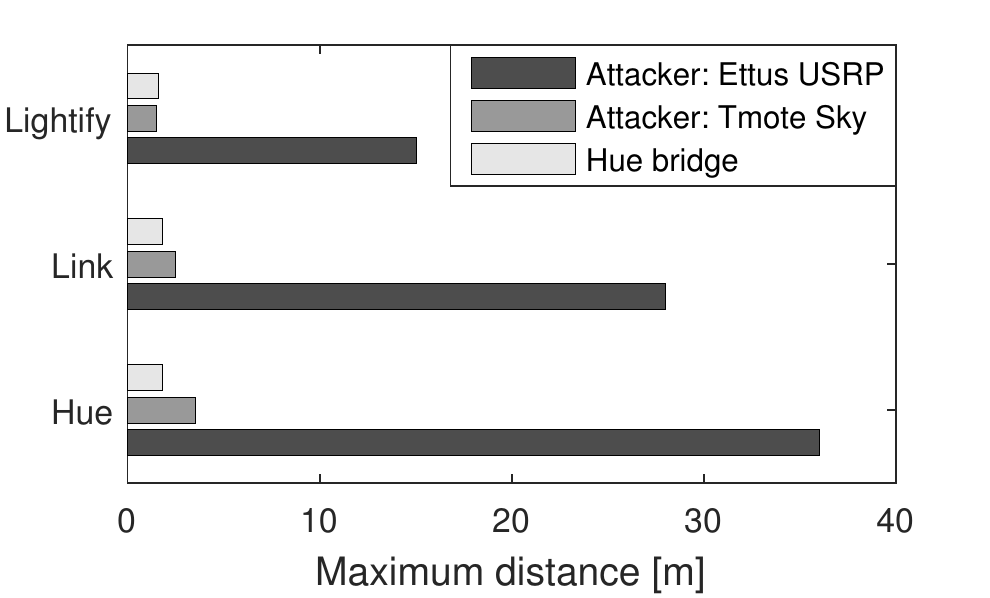}
	\caption{Maximum distances of successfully triggering the touchlink procedure (Hue bridge) or the blink attack (Tmote Sky, Ettus USRP)  on the evaluated connected lighting systems.}
	\label{fig:range}
\end{figure}

After we evaluated the feasibility of the attacks, we analyzed their wireless range. The results are depicted in \Cref{fig:range}.

% Figure: Philips Hue Range 
In the ZLL specification, the manufacturers are advised to limit the wireless range of devices such that only ZLL devices in close proximity are able to perform the ZLL touchlink procedure. This limitation shall be implemented in a way that the received signal strength of an initiator device must be above a certain threshold.
First of all, we measured the maximum distance to successfully perform the touchlink commissioning procedure with a Hue bridge since the Lightify gateway and the Link hub provide no possibility to trigger the touchlink commissioning procedure.
The maximum distances to successfully perform the touchlink commissioning procedure with the Hue bridge is 1.8 meters for a Hue bulb as well as a Link bulb, and 1.6 meters for a Lightify bulb.

Since the touchlink commissioning procedure is intended to require close proximity, we investigated whether the attacks work for longer distances.
Therefore, we evaluated the two radio transceivers, presented in \Cref{sec:analysis:methodology}, regarding the wireless attack range. 

\subsubsection{Moteiv Tmote Sky}
\label{sec:analysis:range:tmote}

The Tmote sky node from Moteiv is suitable for short-distance scenarios. 
Our evaluations shows that we have a wireless range (line-of-sight) of up to 3-4m to attack a Hue bulb, while we need to be within the range of 1-2m (line-of-sight) to successful attack the Lightify bulb. To attack the Link bulb, we must be in the range of 2-3m (line-of-sight). These distances have been measured indoor.

\subsubsection{Ettus USRP B200}
\label{sec:analysis:range:usrp}

\begin{figure}[htb]
        \centering
        \begin{subfigure}[b]{0.23\textwidth}
            \centering
            \includegraphics[width=\textwidth]{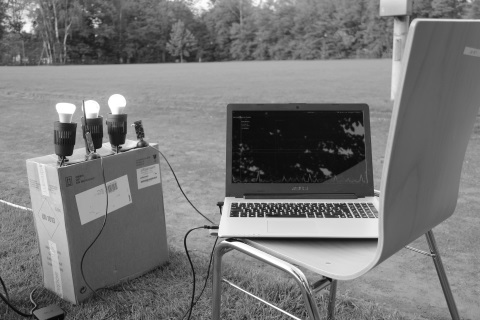}
            \caption[]%
            {{\small Evaluated light bulbs.}}
            \label{fig:outdoor:bulbs}
        \end{subfigure}
       % \hfill
        \begin{subfigure}[b]{0.23\textwidth}
            \centering
            \includegraphics[width=\textwidth]{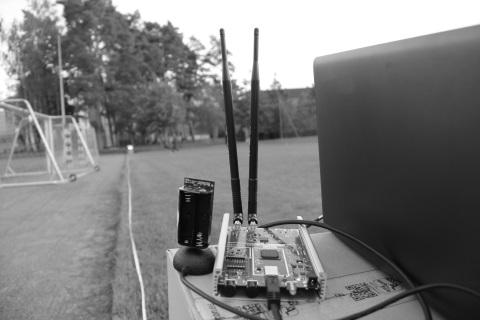}
            \caption[]%
            {{\small Attacker equipment.}}
            \label{fig:outdoor:usrp}
        \end{subfigure}
        \caption{Outdoor testbed to measure the maximum distance and received signal strength of successfully attacking a ZLL-certified light bulb.}
        \label{fig:outdoor}
    \end{figure}

We measured the maximum distance from which we are able to trigger a blink attack. Therefore, we set up an outdoor testbed on a sports ground, in which a line-of-sight between the USRP and the attacked bulb was given. We also recorded the receiving signal strength using the Ubertooth spectrum analyzer \cite{ubertooth}. The outdoor setup is shown in \Cref{fig:outdoor}. 

% Figure: Outdoor testbed

At the USRP, we mounted antennas with 8dB gain (according to the manufacturer).
Our evaluation showed that the three connected lighting systems have different ranges. The maximum distance of the Osram Lightify system is 15 meters, the maximum distance of the GE Link system is 28 meters and the maximum distance of the Philips Hue system is 37 meters.
The distances depend on the noise of the channel as well as the orientation of the bulbs and the antennas of the USRP.  
We experimented with different gain and antenna settings and also with different positions and directions of the bulbs. 
From our measurement results, we estimate that the received signal strength of inter-PAN command frames has to be stronger than -40dBm.

The limitation of range is due to the fact that the USRP did not receive the \emph{scan responses} of the bulbs anymore due to the low transmission power of the bulbs. However, with a parabolic antenna, we assume that the receiving signal strength of the \emph{scan responses} could be amplified in order to increase the maximum distance even further.

%% file: s6_discussion.tex
\section{Discussion}
\label{sec:discussion}

In the first part, we discuss whether the security goals in ZLL-based connected lighting systems are fulfilled. Then, we outline which level of security meets the requirements of connected lighting systems in the second part. In the third part, we distill general security lessons learned for arbitrary Internet-of-Things (IoT) systems.

\subsection{Are Security Goals Satisfied?}
\label{sec:discussion:goals}

Summarizing the related work as well as our novel attacks, we realize that ZLL-based connected lighting systems are insecure against local attackers. 
An attacker can take complete control of any ZLL-certified lighting system since the security weaknesses exist by design.
Also, it is irrelevant whether the targeted ZLL networks have been set up using the classical or the touchlink commissioning procedure.  
The attacks can be performed at any time, in any order and in any number of repeats. 
Furthermore, we found out that close proximity is not required. 
In our evaluation, we successfully attacked the most popular connected lighting system, Philips Hue, from a distance of 36 meters.

Besides insecurities in the ZLL standard, connected lighting systems show a wide variety of other security weaknesses: insufficient command authentication, unprotected hardware debug interfaces, and stored key material in plaintext in the manufacturer apps for mobile devices (cf. \Cref{sec:related}).

To sum up, we state that the security goals for ZLL-based connected lighting systems, presented in \Cref{sec:zigbee:goals}, are not satisfied.

\subsection{Should The Bulbs Be Protected?}
\label{sec:discussion:bulb}

Light bulbs neither gather sensitive information like sensors (which could change in the future) nor put the physical integrity of humans at immediate risk (as opposed, e.g., to malfunctioning medical devices). As a consequence, we need to answer the following questions: Which level of security is needed for light bulbs? What are acceptable risks?

The motivation for attacks on connected lighting systems might not be obvious at the first glance but becomes apparent when considering different deployment areas of these systems.
Although the ZLL standard is originally designed for residential lighting applications, it is currently also deployed in the industrial settings \cite{fortune}.  Based on this recent development, we assume that it will spread also in further domains and therefore the motivation for attacks will increase.
We present three malicious motivations for attackers: to annoy people, to cause financial damage, or to impair health.

In residential lighting applications, an attacker might only annoy or frighten residents by controlling the lights. 
If an attacker can cause annoyance in businesses, however, she can entail financial loss. 
In the hotel business, guests might cancel their rooms resulting in a financial outage. 
Considering companies, the productivity of employees can be limited due to lighting blackouts. 
When lights constitute a central part in work processes, e.g., through directing a workflow, even a complete production standstill could be triggered. 
In case the attackers' motivations is to harm people, she could attack connected lighting systems in medical offices and hospitals. Misconducting lights could interrupt surgeries, or when used for indication of vitals, cause faulty treatments. Also causing epileptic seizures with flickering lights is feasible.

A variety of radio transceivers exists that is capable to perform the attacks presented in \Cref{sec:analysis}. Our equipment, the Ettus USRP, is available for less than \$700 on the US market. 
The attacker must be capable of configuring the radio transceiver and needs an implementation of the attack procedures. Considering an increasing popularity of connected lighting systems and the interest in attack equipment, gray market vendors might develop ready-to-use attack hardware.

 In conclusion, assuming a high motivation of an capable person to attack a connected lighting system, the risk that the system can be controlled by the attacker is high.

\subsection{Lessons Learned for IoT Security}
\label{sec:discussion:lessons}

Learning from the security analysis, we realize that the design of a security architecture for connected lighting systems (and also other IoT devices) is not trivial.
At the first sight, the security of the system seems solid as the specification applies the well-known encryption and authentication scheme AES-CCM*. 
Assuming a shared network key, an attacker without knowledge of this key is not able to decrypt or manipulate the AES-CCM*-encrypted messages.
But the results of our security analysis show that nevertheless an attacker is able to gain full control of the devices by sending valid messages to the bulbs.
In the following, we summarize the lessons learned from the weaknesses of the ZLL lighting systems, which we can also apply to other IoT devices.
During our investigations, we extracted four critical points of failures that weakens the security architecture of the system.

\paragraphsmall{Use of Global Secret.} The first critical point is the trust in the safe-keeping of a pre-shared master key that is shared among multiple manufacturers.
IoT systems set high demands regarding the key management because network keys must be established in very heterogeneous systems, e.g., across different applications and across products of different vendors.
As we see in the case example of ZLL-based connected lighting systems, both commissioning procedures rely on an NDA-protected shared key (used to derive the network key). 
The ZLL link key, used for the classical commissioning, is not leaked yet, but of course, it can happen anytime. That keys are prone to be leaked, even if they are bound by an NDA, can be seen in the example of the ZLL master key.
More suitable solutions do not rely on the secrecy of a master key. 
In the last decade, several key bootstrapping techniques have been proposed and evaluated, e.g.,  Resurrecting Duckling \cite{DBLP:conf/spw/StajanoA99}, which provide the functionality to establish a network key between several devices. Another wireless low-power standard, Bluetooth Low Energy, introduced in version 4.2 a new pairing model that uses elliptic curve cryptography to establish a network key \cite{ble42}.

\paragraphsmall{Fallback Mechanisms.} The second critical point is the implementation of fallback mechanisms in the key management and the network protocol. 
To fulfill the demands of functionality, the ZLL standard deploys fallback solutions in case there exists no pre-shared key material.
The motivation might be to provide as high compatibility as possible. In the case of ZLL, to allow the communication with other ZigBee (but non-ZLL) devices.
In general, a trade-off between security and functionality is needed in IoT devices. User might be rather willing to lower the security expectations compared to having incompatible devices that cannot be added to the home network due to missing key material.
The design of a security architecture must be simple and straight forward such that there is no need for fallback mechanisms. 
Furthermore, network protocols should be designed such that only authorized commands overrule existing settings, e.g., neither to join a bulb from one to another network nor the reset-to-factory-state should be possible without authorization. 

\paragraphsmall{Tamper-Resistent Hardware.} The third critical point of failure is the missing of tamper resistant hardware in IoT devices. There are numerous examples of extracting firmware and cryptographic material from IoT devices exploiting unprotected debug interfaces \cite{chapman2014, heres2014hack, wineberg}. 
In recent years, a huge effort was put to protect other technologies like smart cards, mobile phones and computers against tampering attacks.
Because the time-to-market and costs supposedly constrain the security development of IoT devices, tamper-resistance is not a priority. 
From our point of view, the attacker model for IoT devices should  consider hardware tampering attacks as well as the extraction and reverse-engineering of firmware images.

\paragraphsmall{Reliance on Signal Strength.} The fourth critical point of failure is managing access to IoT networks based on (simple) physical properties like signal strength.
Using a software-defined radio platform, an attacker can easily adjust the transmit power to higher values as expected by certified products.
In this way, security measures can be circumvented. As shown in \Cref{sec:analysis}, in the case of the Philips Hue lighting system, we are able to send valid messages from a distance of over 30 meters, while the manufacturer implemented a limitation aiming a wireless range of 2-3 meters.

%% file: s9_conclusion.tex
\section{Conclusion}
\label{sec:conclusion}

Since the introduction to the market in 2012, connected lighting systems showed a wide variety of insecurities, from unprotected debug interfaces to fallback mechanisms in the network protocol and unauthenticated command messages.
We presented a comprehensive overview of security research regarding connected lighting systems, and performed a security analysis of the most popular standard of these systems, ZigBee Light Link.  
We disclosed further attacks in the touchlink commissioning procedure and evaluated their impact using our implemented penetration testing framework. Learning from the security pitfalls of ZLL-based connected lighting systems, we suggested improvements for securing connected lighting systems and derived lessons for other IoT systems.

Connected lighting systems based on the ZLL standard need improvements of security measures. The coming ZigBee 3.0 standard is announced to replace the ZLL profile.  Future investigations are required to determine whether the security of connected lighting systems is improved with ZigBee 3.0, or not. 
Also the ZigBee firmware update mechanisms provide an attractive target to evaluate if spreading malware on IoT devices is feasible.